\documentclass{article}

\usepackage{microtype}
\usepackage{graphicx}
\usepackage{subcaption}
\usepackage{booktabs} 
\usepackage[most]{tcolorbox}
\tcbuselibrary{listings}

\usepackage[hidelinks]{hyperref}

\usepackage[preprint]{icml2026}

\usepackage{amsmath}
\usepackage{amssymb}
\usepackage{mathtools}
\usepackage{amsthm}
\usepackage{tabularx}
\usepackage{enumitem}
\usepackage[capitalize,noabbrev]{cleveref}

\theoremstyle{plain}

\theoremstyle{definition}

\theoremstyle{remark}

\usepackage[textsize=tiny]{todonotes}

\icmltitlerunning{AlphaLogics for Market Logic-Driven Factor Generation}

\begin{document}

\twocolumn[
  \icmltitle{AlphaLogics: A Market Logic-Driven Multi-Agent System for \\Scalable and Interpretable Alpha Factor Generation}



  \begin{icmlauthorlist}
    \icmlauthor{Zhangyuhua Weng}{yyy}
    \icmlauthor{Shengli Zhang}{yyy}
    \icmlauthor{Taotao Wang}{yyy}
    \icmlauthor{Yihan Xia}{yyy}
  \end{icmlauthorlist}

  \vspace{0.1em}
  \begin{center}
    \small College of Electronic and Information Engineering, Shenzhen University\\
    \small \texttt{2410043007@mails.szu.edu.cn, zsl@szu.edu.cn, ttwang@szu.edu.cn, xiayihan2023@email.szu.edu.cn}
  \end{center}

  \icmlaffiliation{yyy}{College of Electronic and Information Engineering, Shenzhen University}

  \icmlcorrespondingauthor{Zhangyuhua Weng}{2410043007@mails.szu.edu.cn}
  \icmlcorrespondingauthor{Shengli Zhang}{zsl@szu.edu.cn}
  \icmlcorrespondingauthor{Taotao Wang}{ttwang@szu.edu.cn}
  \icmlcorrespondingauthor{Yihan Xia}{xiayihan2023@email.szu.edu.cn}

  \icmlkeywords{Factor Investing, Large Language Models, Multi-Agent Systems, Quantitative Finance, Market Logic}

  \vskip 0.3in
]



\makeatletter
\renewcommand{\printAffiliationsAndNotice}[1]{\global\icml@noticeprintedtrue}
\makeatother

\printAffiliationsAndNotice{}  

\begin{abstract}
Factor investing is ultimately grounded in market logic—the latent mechanism behind observed alpha factors that explains why they should persist across assets and regimes. However, recent factor mining prioritizes factor discovery over logic discovery, producing complex alpha factors with unclear rationale, while market logic remains largely handcrafted and difficult to scale. To address this challenge, we propose AlphaLogics, a market logic-driven multi-agent system for factor mining. AlphaLogics consists of three key components: (i) Market Logic Mining: reverse-extracting market logic from historical factor libraries to construct an initial market logic library; (ii) Guided Factor Generation: using given market logics (generated in i\&iii) to guide new factors generation and
optimization with backtesting feedback; and (iii) Market Logic Generation: generating new market logics conditioned on the initial market logic library, and refining each market logic by aggregating the backtest outcomes of its guided factors, continuously refreshing the library. Experiments on CSI 500 and S\&P 500 show that AlphaLogics consistently improves predictive metrics and risk-adjusted returns over representative baselines, while producing a market logic library that remains empirically useful for guiding further factor discovery.
\end{abstract}

\section{Introduction}

With the rapid digitalization of financial markets, quantitative investing increasingly relies on alpha factors as core building blocks that shape return stability and risk exposure \cite{gu2020empirical}. Classical factor design is grounded in market logic, offering interpretability but limited scalability, whereas market data-driven mining scales efficiently yet often lacks economic grounding and robustness \cite{zhang2020autoalpha}. Bridging this trade-off—scalable factor generation with interpretable market logic and cross-cycle robustness—remains a central challenge in quantitative finance \cite{zhang2020autoalpha}.

Traditional factor research starts from market logic, distilling economic regularities into structured mathematical forms \cite{fama2015five}; Historical factor libraries such as Alpha191 embed interpretable reasoning (e.g., momentum, reversal, volatility, and price--volume interaction) and show practical relevance \cite{guotai191alphafactor}. However, this paradigm is slow and difficult to scale because discovery depends on expert iteration and extensive cross-cycle validation \cite{cui2021alphaevolve}. In contrast, Machine learning methods can generate massive candidate pools—ranging from widely used composite factors in Alpha101 to large-scale automated frameworks \cite{kakushadze2016,zhang2020autoalpha,ren2024alphamining,shi2025alphaforge}—but the resulting factors are often hard to interpret due to missing theoretical or behavioral constraints \cite{tatsat2025beyond,rudin2019why,tong2024ploutos}.

Large language models (LLMs) offer a promising path to inject domain knowledge and structured reasoning into factor generation \cite{nie2024survey_llm_finance,ding2023integrating}. Existing systems such as Alpha-GPT and AlphaAgent demonstrate that natural-language ideas and semantic constraints can guide factor construction and mitigate performance decay \cite{wang2025alpha_gpt,tang2025alphaagent}. Nevertheless, prior LLM-based frameworks predominantly optimize factors, rather than treating market logic itself as an explicit, verifiable, and optimizable research object.

In scientific modeling, explicit hypotheses serve as structural priors that constrain the search space and enhance interpretability. Quantitative investing implicitly relies on this mechanism as well, yet the acquisition of high-quality market logic has long been limited by expert experience and heuristic exploration.

To address this gap, we propose AlphaLogics, a market logic-driven multi-agent system for factor mining, which treats market logic as an explicit and iteratively improvable object. AlphaLogics consists of three key components: (i) \textbf{Market Logic Mining}. This Stage reverse-extracts market logic from historical public factor libraries—Alpha101 \cite{kakushadze2016}, Alpha191 \cite{guotai191alphafactor}, Alpha158 \cite{qlib_alpha158}, and Alpha360 \cite{qlib_alpha360}—motivated by their empirical robustness across markets \cite{gu2020empirical,afzal2023volatility,hibbeln2025model_validation} to construct an initial market logic library. (ii) \textbf{Guided Factor Generation}. Using new market logics (\emph{generated in i\&iii}) as guidance, this Stage generates candidate factors and optimizes them with backtesting feedback. (iii) \textbf{Market Logic Generation}. This Stage generates new market logics conditioned on the initial market logic library, and refines each market logic by aggregating the backtest performances of its guided factors, continuously refreshing the market logic library. In this work, we focus on technical market logic derived from OHLCV and price--volume patterns, and do not model fundamental or accounting-based signals.

We evaluate the AlphaLogics on CSI 500 (China) and S\&P 500 (U.S.) from January 2021 to December 2024(On the held-out test period); after trading costs, it achieves annualized excess returns of 16.72\% (IR = 1.5266) and 13.75\% (IR = 1.2658), respectively, and further experiments show that iterative market-logic optimization improves the quality of logic-guided factors over time.

The main contributions of this work are summarized as follows:
\begin{enumerate}
  \item \textbf{Construction of a market logic library:} We extract underlying market logic from historical factor libraries, building a comprehensive repository of reliable market logics to support factor engineering.
  \item \textbf{Automated generation and optimization of market logic:} AlphaLogics mitigates the reliance on human-generated market logic by generating and iteratively refining new market logic, enhancing both scalability and quality.
  \item \textbf{Extensive multi-market validation:} Empirical experiments across multiple markets show that factors generated under AlphaLogics achieve superior performance compared to existing methods, while the associated market logic remains more transparent and interpretable.
\end{enumerate}

\begin{figure}[t]
  \centering
  \includegraphics[width=0.96\columnwidth]{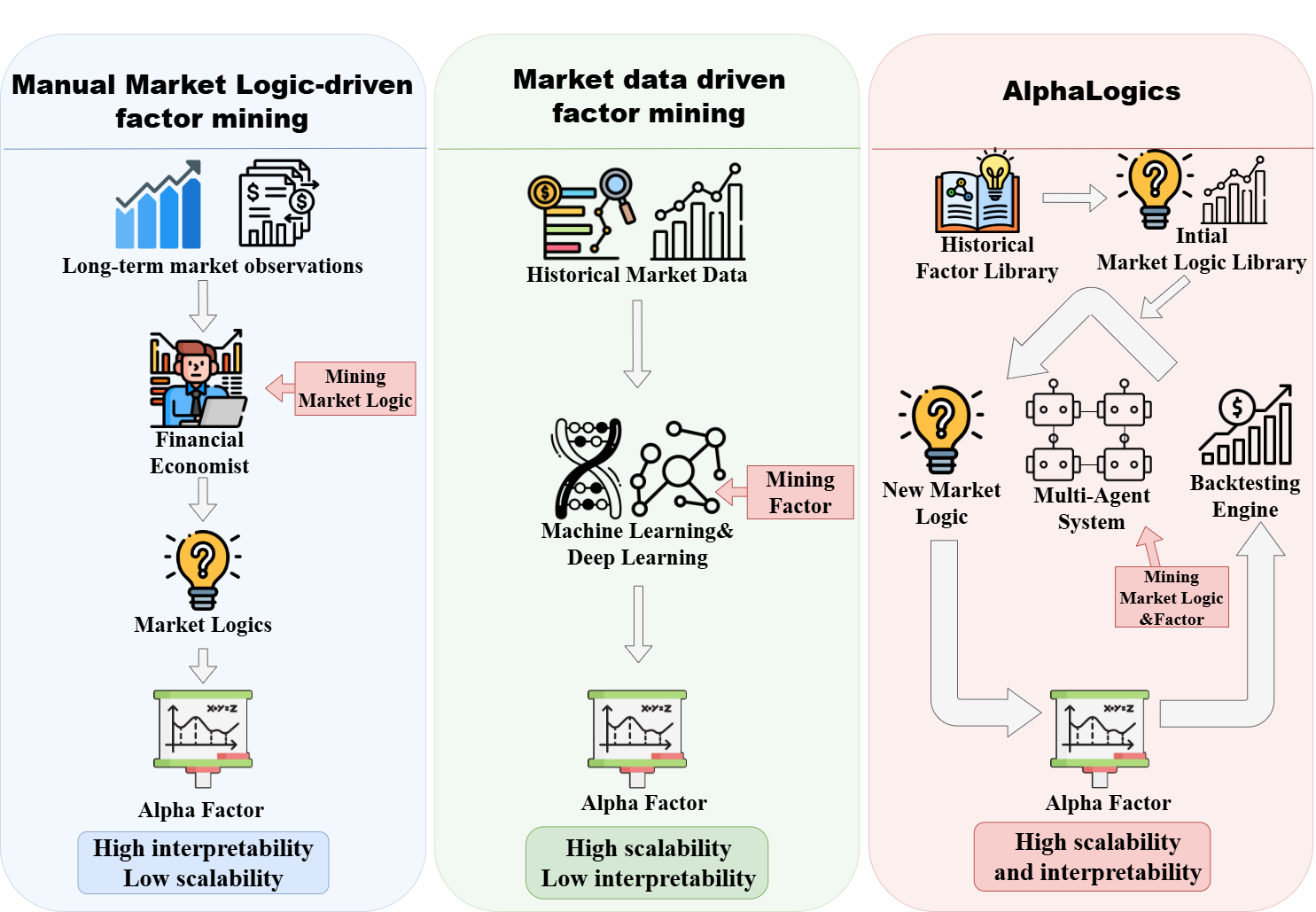}
  \caption{\small
  Comparison of advantages and disadvantages of alpha factor mining methods. Manual Market Logic-driven factor mining (left) offers high interpretability but lacks scalability; Market data driven factor mining (middle) scales well but often lacks interpretability; Our method (right) offers high interpretability and scalability.
  }
  \label{fig:alpha_evolution}
\end{figure}


\section{Related Work}

Factor construction sits at the core of quantitative investing and has evolved with asset pricing theory and AI. We review three streams: Manual Market Logic-driven design, Market Data-Driven modeling, and LLM-driven factor construction, highlighting their trade-offs and the gap our work targets.

\subsection{Manual Market Logic-Driven Factor Construction}

Classical factor models derive economically interpretable pricing factors from observables such as fundamentals and market behavior. The Fama--French three-factor model \cite{fama1993common} and its five- and six-factor extensions \cite{fama2015five,fama2018choosing} embed explicit market logic around size, value, profitability, investment, and momentum. These approaches offer clear economic narratives but rely on expert design and long validation cycles, limiting scalability and responsiveness in high-dimensional markets.

\subsection{Market Data-Driven Factor Construction}

Machine learning enables scalable, data-driven factor construction \cite{gu2020empirical,ye2024from_factor_models}. Nonlinear models and automatic feature extraction capture complex signals from price, volume, text, and microstructure \cite{gu2020empirical,liu2025deep_conditional,feng2024deep_learning_character}. However, these factors often lack explicit market logic, weakening interpretability and theoretical grounding and exacerbating regime instability and overfitting concerns \cite{liao2025uncertainty_ml}.

\subsection{LLM-Driven Factor Construction}

LLM-based frameworks largely focus on automating factor generation and refinement via generation–backtesting loops \cite{kou2024automate_strategy,duan2025factormad,tang2025alphaagent}, while leaving market logic implicit and manually crafted rather than generated and evolved—making it time-consuming to build and hard to scale.

To bridge this gap, we propose AlphaLogics, which treats market logic as an explicit, optimizable object to combine scalability with interpretability.

\section{Methodology}

\begin{figure*}[t]
    \centering
    \includegraphics[width=0.88\textwidth]{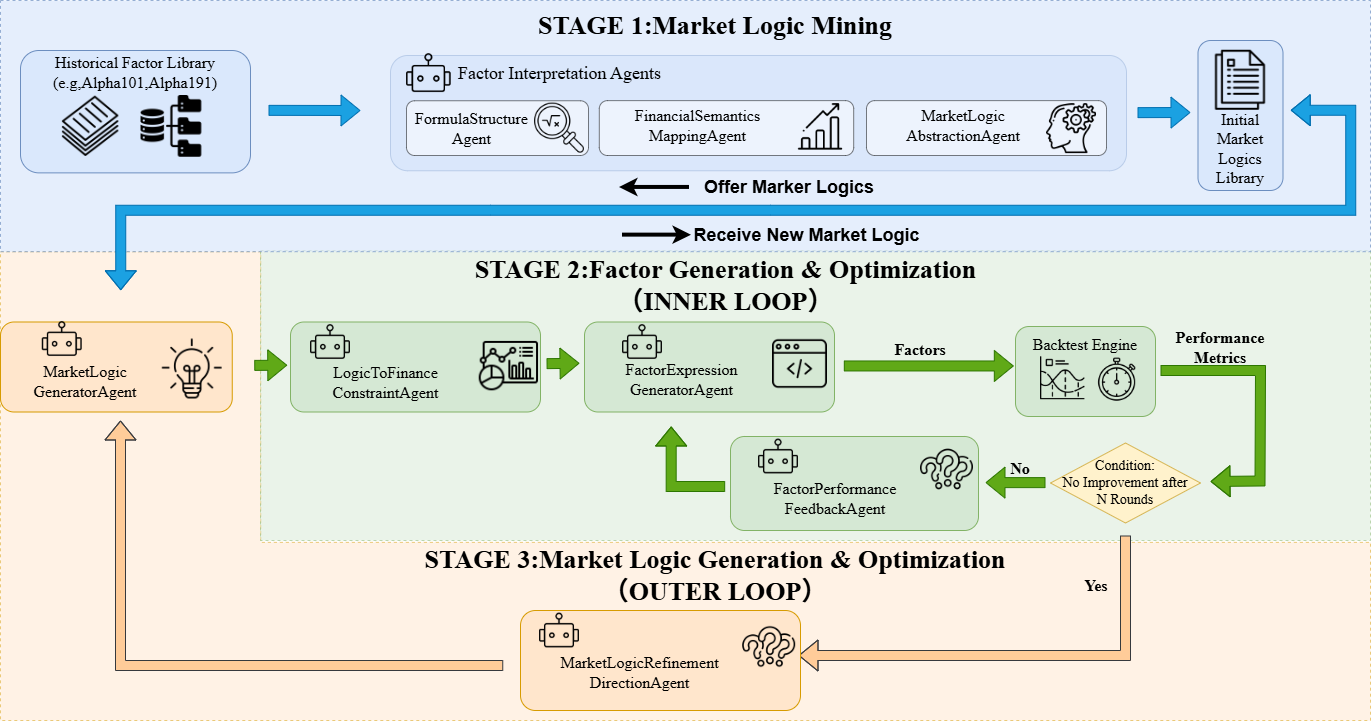}
    \caption{
    The autonomous workflow of \textbf{AlphaLogics}. The framework operates in three stages:
    (1) \textit{Market Logic Mining} extracts latent market logic from historical factor libraries;
    (2) \textit{Guided Factor Generation} guides new factors generation and optimization with backtesting feedback;
    (3) \textit{Market Logic Generation} generates new market logics conditioned on the initial market logic library, and refining each market logic by aggregating the backtest outcomes of its guided factors, continuously refreshing the library.
    }
    \label{fig:framework}
\end{figure*}

Figure~\ref{fig:framework} summarizes AlphaLogics in three stages: (1) \textbf{Market Logic Mining} to build an initial logic library from historical factor libraries, (2) \textbf{Factor Generation} to generate and refine factors with backtesting feedback under given market logics, and (3) \textbf{Market Logic Generation} to generate new market logics conditioned on the initial market logic library, and refining each market logic by aggregating the backtest outcomes of its guided factors, continuously refreshing the library.

\subsection{Market Logic Mining}
\label{subsec:logic_mining}
We extract market logic from historical factor libraries via a structured multi-agent interpretation pipeline.

We mine latent logic from Alpha101 \cite{kakushadze2016}, Alpha191 \cite{guotai191alphafactor}, Alpha158 \cite{qlib_alpha158}, and Alpha360 \cite{qlib_alpha360}, using three agents to abstract factor expressions from ``mathematical formula $\rightarrow$ financial meaning $\rightarrow$ market logic''.
We represent a single market logic as the following pair:

\begin{equation}
H = \langle \mathcal{C}, \mathcal{B} \rangle
\end{equation}

where $\mathcal{C}$ is a conjunction/disjunction of predicates $c_i=(v, \mathrm{op}, \theta, w)$ over market variables, and $\mathcal{B}=(y,d,h)$ specifies target, direction, and horizon. We distinguish the human-readable $H$, structured $H^{\text{struct}}$, and executable constraints $\Gamma=\textsc{Compile}(H^{\text{struct}})$, which constrain variables, operator families, parameter ranges, and sign/direction. Appendix~\ref{app:compile} provides the schema and a \textsc{Compile} example.Here $H$ is human-readable and specifies only coarse $\mathcal{C}$ and $\mathcal{B}$. The structured instantiation into $H^{\text{struct}}$ and compilation into $\Gamma$ are performed later by the LogicToFinanceConstraintAgent (Section~3.2).


\textbf{FormulaStructureAgent.} The FormulaStructureAgent takes a factor's mathematical formula as input and focuses on analyzing its operational structure and formal properties. The output is a structured description of the factor's computational logic without introducing financial interpretation.

\textbf{FinancialSemanticsMappingAgent.} After obtaining the structural logic, the FinancialSemanticsMappingAgent maps mathematical operations into canonical financial and behavioral-finance semantics.

\textbf{MarketLogicAbstractionAgent.} Finally, the MarketLogicAbstractionAgent abstracts financial semantics into explicit market logic, expressed in natural language. The output is a human-readable market logic $H$ with explicit $\mathcal{C}, \mathcal{B}$ semantics, which is later canonicalized into $H^{\text{struct}}$.

The example below illustrates factor-to-market logic extraction.
\begin{center}
\begin{tcolorbox}[width=0.92\linewidth, colback=white, colframe=black, breakable, boxrule=0.5pt]
\footnotesize
\textbf{Example Factor:}
\[
\begin{split}
f = -\mathrm{TS\_CORR}(&\mathrm{RANK}(\mathrm{open}),\\
                         &\mathrm{RANK}(\mathrm{volume}),\\
                         &10)
\end{split}
\]
\textbf{FormulaStructureAgent} : Identifies cross-sectional ranking on price and volume, followed by 10-day rolling correlation with sign reversal.

\textbf{FinancialSemanticsMappingAgent} : Maps price to initial reaction and sentiment; maps volume to participation and confirmation; correlation measures persistent price–volume co-movement.

\textbf{MarketLogicAbstractionAgent} : Extracts Market Logic: ``($\mathcal{C}$) Price movements lacking volume confirmation are more likely driven by short-lived sentiment and mean-revert; ($\mathcal{B}$) thus the expected short-horizon forward return is negative, while strong volume confirmation indicates information assimilation and trend continuation.''

\end{tcolorbox}
\end{center}

This step converts historical factors into a scalable, interpretable logic library $\mathcal{H}_{\text{init}}$.

\subsection{Factor Generation \& Optimization}
We search for factors guided by new market logics, using backtesting feedback for optimization.

Formally, a single technical factor can be expressed as:
\begin{equation}
f_t = \mathcal{O}(x_{t-k:t})
\end{equation}
where $x_{t-k:t}$ is the lookback window and $\mathcal{O}$ is an operator composition.

Factors are generated from a fixed DSL with parameterized operators (Appendix~\ref{subsec:app_factor_dsl}).

Algorithm~\ref{alg:inner-loop} fixes $H$, canonicalizes it into $H^{\text{struct}}$, compiles constraints $\Gamma$, and refines factors within $\Gamma$ using backtesting feedback, with both canonicalization and compilation handled by the LogicToFinanceConstraintAgent. Selection and early stopping use $J(R_{\text{val}})$ on $D_{\text{val}}$ only. When improvements stop, we summarize per-logic performance (at minimum the best validation metrics) for the outer loop. Agents are listed below. For brevity in Algorithms~\ref{alg:inner-loop} and \ref{alg:outer-loop}, we use abbreviated agent names: L2FC-Agent (LogicToFinanceConstraintAgent), FEG-Agent (FactorExpressionGeneratorAgent), FPF-Agent (FactorPerformanceFeedbackAgent), MLGA-Agent (MarketLogicGeneratorAgent), and MLRD-Agent (MarketLogicRefinementDirectionAgent).

\begin{algorithm}[t]
\footnotesize
\caption{Inner Loop: Factor Optimization under Fixed Market Logic $H$}
\label{alg:inner-loop}
\begin{algorithmic}
\STATE \textbf{Input:} \parbox[t]{0.86\linewidth}{
  $H$ (human-readable market logic, generated by MLGA-Agent),\\
  $D_{\text{train}}/D_{\text{val}}$ (train/validation splits),\\
  objective $J$ (scalar on validation metrics, e.g., IR), early-stop $T_{\text{early}}$
}
  \STATE {\bfseries Output:} $E^{\text{Logic}}$ (logic-level evidence summary), $R_{\text{val}}^{\star}$ (best validation metrics under $H$)
  \STATE $c \leftarrow 0$; $R_{\text{val}}^{\star} \leftarrow \textsc{None}$
  \STATE $fb^{\text{factor}} \leftarrow \textsc{None}$
  \STATE $(H^{\text{struct}}, \Gamma) \leftarrow \textsc{L2FC-Agent}(H)$
  \WHILE{$c < T_{\text{early}}$}
    \STATE $F \leftarrow \textsc{FEG-Agent}(\Gamma \mid fb^{\text{factor}})$
    \STATE $(R_{\text{train}}, R_{\text{val}}) \leftarrow \textsc{BacktestEngine}(F; D_{\text{train}}, D_{\text{val}})$
    \IF{$R_{\text{val}}^{\star} = \textsc{None}$ \textbf{or} $J(R_{\text{val}}) > J(R_{\text{val}}^{\star})$}
      \STATE $R_{\text{val}}^{\star} \leftarrow R_{\text{val}}$; $c \leftarrow 0$
    \ELSE
      \STATE $c \leftarrow c + 1$
    \ENDIF
    \IF{$c < T_{\text{early}}$}
      \STATE $fb^{\text{factor}} \leftarrow \textsc{FPF-Agent}(H^{\text{struct}}, F, R_{\text{val}})$
    \ENDIF
  \ENDWHILE
  \STATE $E^{\text{Logic}} \leftarrow \langle H^{\text{struct}}, R_{\text{val}}^{\star} \rangle$
  \RETURN $E^{\text{Logic}}, R_{\text{val}}^{\star}$
\end{algorithmic}
\end{algorithm}

\textbf{LogicToFinanceConstraintAgent:} Transforms the human-readable market logic $H$ produced by the MarketLogicAbstractionAgent into a normalized structured form $H^{\text{struct}}$ by standardizing variable names and filling required fields without changing semantics, and then implements $\Gamma = \textsc{Compile}(H^{\text{struct}})$ to map structured market logic to executable constraints over variables, operator families, parameter ranges, and sign/direction consistency implied by $\mathcal{B}$. The goal of this stage is to convert the high-level market logic into a constraint description with financial semantic consistency.

\textbf{FactorExpressionGeneratorAgent:} Under the financial semantics and operational constraints provided by $\Gamma$, the FactorExpressionGeneratorAgent generates mathematical factor expressions that satisfy the market logical semantics. The generation process involves the selection of basic operators, time windows, parameter values, and operator combination methods, enabling the factors to be directly calculated and back-tested in the quantitative trading system.

\textbf{Backtest Engine:} Given factors $F$ generated by the FactorExpressionGeneratorAgent and historical data split into $D_{\text{train}}$, $D_{\text{val}}$, and $D_{\text{test}}$, the Backtest Engine runs training and validation backtests and returns $(R_{\text{train}}, R_{\text{val}})$. Here $R_{\text{train}}, R_{\text{val}}$ are the corresponding sets of backtest metrics (e.g., IC/IR/AR/MDD). The final backtest results are reported on $D_{\text{test}}$.

\textbf{FactorPerformanceFeedbackAgent:} Because LLM generation is stochastic, factors produced under the same logic can vary. The agent maintains a buffer of the most recent $M$ candidates and their $D_{\text{val}}$ metrics under a fixed $\Gamma$, performs cross-candidate comparison, and returns structured feedback to guide the next generation. We set $M$ to the per-logic candidate budget and reset the buffer when switching to a new market logic.

If factors' performance on $D_{\text{val}}$ \textup{(}$J(R_{\text{val}})$\textup{)} stops improving, the Inner Loop terminates and returns a logic-level evidence summary $E^{\text{Logic}}$ (at minimum including $R_{\text{val}}^{\star}$) for the outer loop (Section~\ref{subsec:logic_optimization}).

\subsection{Market Logic Generation \& Optimization}
\label{subsec:logic_optimization}
We optimize market logic itself when factor performance under a fixed logic stops improving.

When successive rounds under a fixed logic stop improving, the testable prediction structure of that logic is likely saturated rather than the factor search space being too small.

Algorithm~\ref{alg:outer-loop} generates and refines market logic based on accumulated performance from the Inner Loop. Each newly generated market logic is appended to the market logic library, which grows over rounds and becomes part of the conditioning context for subsequent generation. Each logic is treated as a testable objection, and we keep the best $H^{\star}$ by maximizing $J(R_{\text{val}}^{\star})$ on $D_{\text{val}}$. The outer loop calls the MarketLogicRefinementDirectionAgent after each inner-loop run to aggregate historical evidence and propose refinement directions. 

\begin{algorithm}[t]
\small
\caption{Outer Loop: Market Logic Optimization}
\label{alg:outer-loop}
\begin{algorithmic}
  \STATE {\bfseries Input:} $\mathcal{H}_{\text{init}}$: Initial logic library; $D_{\text{train}}$, $D_{\text{val}}$
  \STATE $T$: Max optimization attempts; $J(\cdot)$: scalar objective on $D_{\text{val}}$
  \STATE {\bfseries Output:} $H^{\star}$: Optimized market logic
  \STATE $\mathcal{H}_{\text{lib}} \leftarrow \mathcal{H}_{\text{init}}$
  \STATE $H_{\text{current}} \leftarrow \textsc{MLGA-Agent}(\mathcal{H}_{\text{lib}})$
  \STATE $\mathcal{H}_{\text{lib}}.\text{append}(H_{\text{current}})$
  \STATE $H_{\text{hist}} \leftarrow [H_{\text{current}}]$; $\mathcal{E}_{\text{hist}} \leftarrow [\ ]$; $fb^{\text{Logic}}_{\text{hist}} \leftarrow [\ ]$
  \STATE $H_{\text{best}} \leftarrow H_{\text{current}}$; $\mathcal{E}_{\text{best}} \leftarrow \textsc{None}$; $t \leftarrow 0$
  \WHILE{$t < T$}
    \STATE $(E^{\text{Logic}}, R_{\text{val}}^{\star}) \leftarrow \textsc{InnerLoop}(H_{\text{current}}; D_{\text{train}}, D_{\text{val}})$
    \STATE Update history: $\mathcal{E}_{\text{hist}}.\text{append}(E^{\text{Logic}})$
    \STATE $fb^{\text{Logic}} \leftarrow \textsc{MLRD-Agent}(H_{\text{current}}, H_{\text{hist}}, \mathcal{E}_{\text{hist}}, fb^{\text{Logic}}_{\text{hist}})$
    \STATE $fb^{\text{Logic}}_{\text{hist}}.\text{append}(fb^{\text{Logic}})$
    \IF{$\mathcal{E}_{\text{best}} = \textsc{None}$ \textbf{or} $J(R_{\text{val}}^{\star}) > J(\mathcal{E}_{\text{best}})$}
      \STATE $H_{\text{best}} \leftarrow H_{\text{current}}$; $\mathcal{E}_{\text{best}} \leftarrow R_{\text{val}}^{\star}$
    \ENDIF
    \STATE $H_{\text{new}} \leftarrow \textsc{MLGA-Agent}(\mathcal{H}_{\text{lib}}, H_{\text{current}}, H_{\text{hist}}, \mathcal{E}_{\text{hist}}, fb^{\text{Logic}}_{\text{hist}})$
    \STATE $H_{\text{hist}}.\text{append}(H_{\text{new}})$
    \STATE $\mathcal{H}_{\text{lib}}.\text{append}(H_{\text{new}})$
    \STATE $H_{\text{current}} \leftarrow H_{\text{new}}$; $t \leftarrow t + 1$
  \ENDWHILE
  \RETURN $H^{\star} \leftarrow H_{\text{best}}$
\end{algorithmic}
\end{algorithm}

\textbf{MarketLogicGeneratorAgent:}  Generates human-readable market logic $H = \langle \mathcal{C}, \mathcal{B} \rangle$. In the first round it conditions only on the initial market logic library $\mathcal{H}_{\text{init}}$; in later rounds it conditions on the expanded library together with the current market logic $\mathcal{H}_{\text{current}}$, market logic history $\mathcal{H}_{\text{hist}}$, and accumulated feedback to propose refinements. Each newly generated market logic is appended to the library and becomes part of the conditioning context in subsequent rounds. The role of this Agent is not to randomly generate text descriptions, but to diverge based on the extracted market logics, ensuring that the new market logics still correspond to market mechanisms that can be quantitatively characterized.

\textbf{MarketLogicRefinementDirectionAgent:} The core component of the Outer Loop is the MarketLogicRefinementDirectionAgent, which is invoked after each inner-loop run to act directly on market logic rather than fine-tuning factor expressions. This agent takes the current market logic $\mathcal{H}_{\text{current}}$, the historical market logic set $\mathcal{H}_{\text{hist}}$, the accumulated logic evidence history $\mathcal{E}_{\text{hist}}$ returned by the Inner Loop, and the market logic-level feedback history $fb^{\text{Logic}}_{\text{hist}}$ as inputs. It summarizes and reflects the market logic from a cross-factor perspective and provides refinement suggestions. Specifically, this agent will comprehensively analyze the following information: (1) The performance distribution of different factors in time, market conditions, and risk dimensions under this market logic; (2) Components in the market logic that may be too broad, vague, or mismatched with the actual market structure.

Based on the refinement suggestions, the MarketLogicGeneratorAgent refines or restructures the market logic and feeds it back into the next inner-loop round.

\section{Experiments}

\subsection{Experimental Setup}
\label{subsec:setup}

We conduct backtesting in Qlib \cite{yang2020qlib} on CSI 500 (China A-share) and S\&P 500 (U.S.) using OHLCV data from Baostock \cite{baostock2024} and Yahoo Finance \cite{ranaroussi2024yfinance}. We filter stocks with fewer than 100 trading days and report metrics in Appendix~\ref{subsec:app_metrics}. The data split is train (2015.01--2019.12), validation (2020.01--2020.12), and test (2021.01--2024.12).

\textbf{No-leakage and unified evaluation protocol.} We strictly split $D_{\text{train}}$, $D_{\text{val}}$, and $D_{\text{test}}$. All optimization signals (early stopping, factor/logic selection, and feedback) use $D_{\text{val}}$ only; $D_{\text{train}}$ is used only for fitting LightGBM and computing factor values. The final backtest results are reported on $D_{\text{test}}$. All baselines share the same stock universe, time split, OHLCV inputs, top-outside rule, and cost model, and output a cross-sectional score per date (LightGBM on base plus generated factors for factor methods; model predictions for deep baselines), so differences come from signal construction.

AlphaLogics use four base factors (intraday return, daily return, 20-day relative volume, normalized daily range) and concatenate them with generated factors to train LightGBM \cite{ke2017lightgbm}. Features and returns are cross-sectionally Z-scored. For each market logic, we train LightGBM on $D_{\text{train}}$, evaluate a predefined selection objective $J$ (e.g., validation IR, IC, ICIR, AR, and MDD) on $D_{\text{val}}$ for early stopping and model selection. After training the model on $D_{\text{train}} \cup D_{\text{val}}$, we run a final backtest on $D_{\text{test}}$ and report the resulting performance metrics. Backtesting uses a top-outside strategy (select the 50 top-ranked stocks based on the predicted returns and exclude the 5 lowest-ranked stocks.). Transaction costs are included: CSI 500 buy 0.0005/sell 0.0015; S\&P 500 buy 0.0000/sell 0.0005.The Inner Loop uses an early-stopping threshold of $T_{\text{early}} = 3$.

For agent-based and LLM-based baselines, we align budgets across markets and methods, fixing the number of trials, evolution rounds, and candidate factors per round as described earlier. And all methods are evaluated with the same top-outside strategy. This setup isolates the impact of market logic guidance rather than differences in evaluation protocol.

\subsection{Benchmark Results on CSI 500 and S\&P 500}
We compare AlphaLogics with representative baselines on predictive and portfolio metrics across both markets.

We include deep time-series models, tree-based models, professional quant models, LLM baselines, and multi-agent factor systems (Table~\ref{tab:main_results}). For agent-based baselines (RD-Agent, AlphaAgent) and AlphaLogics, we align budgets: each run uses 20 trials with 5 evolution rounds using GPT-3.5-turbo \cite{ouyang2022training}, with identical caps on LLM calls and candidate factors per round. We also include O3-mini and Deepseek-V3.1 as direct factor generators (no market logic constraints), using the same DSL and data split; results are averaged over 20 independent trials.

\begin{table*}[t]
  \caption{Comparison of factor performance on CSI 500 and S\&P 500 over the held-out test period (2021.01--2024.12), using a fixed split: train (2015.01--2019.12), validation (2020.01--2020.12), and test (2021.01--2024.12). \textbf{Bold} indicates the best performance among all methods for each metric within the same market.}
  \label{tab:main_results}
  \centering
  \small
  \setlength{\tabcolsep}{4pt}
  \begin{tabular}{lccccc ccccc}
    \toprule
    \textbf{Method} & \multicolumn{5}{c}{\textbf{CSI 500 (2021.01--2024.12)}} & \multicolumn{5}{c}{\textbf{S\&P 500 (2021.01--2024.12)}} \\
    \cmidrule(lr){2-6} \cmidrule(lr){7-11}
     & IC & ICIR & AR & IR & MDD & IC & ICIR & AR & IR & MDD \\
    \midrule
    LSTM \cite{graves2012lstm} & 0.0162 & 0.1173 & 6.33\% & 0.8494 & -10.45\% & 0.0032 & 0.0177 & -7.29\% & -0.9732 & -21.91\% \\
    Transformer \cite{vaswani2017attention} & 0.0150 & 0.1234 & 4.03\% & 0.3650 & -22.85\% & 0.0014 & 0.0108 & 0.82\% & 0.0952 & -17.71\% \\
    LightGBM \cite{ke2017lightgbm} & 0.0116 & 0.0972 & 0.88\% & 0.0782 & -24.37\% & -0.0014 & -0.0131 & -1.23\% & -0.1517 & -28.28\% \\
    TRA \cite{lin2021temporal} & 0.0199 & 0.1765 & 1.60\% & 0.1467 & -26.11\% & 0.0030 & 0.0204 & -3.92\% & -0.4541 & -34.15\% \\
    GRU \cite{chung2014gru} & 0.0115 & 0.0994 & -1.71\% & -0.1733 & -22.05\% & 0.0050 & 0.0297 & -1.14\% & -0.0970 & -32.10\% \\
    XGBoost \cite{chen2016xgboost} & 0.0122 & 0.1066 & 5.37\% & 0.4372 & -26.33\% & -0.0028 & -0.0056 & -0.65\% & -0.0936 & -22.63\% \\
    MLP \cite{taud2017mlp} & 0.0115 & 0.0975 & 5.42\% & 0.4671 & -23.34\% & -0.0007 & -0.0056 & -1.09\% & -0.1432 & -23.33\% \\
    O3-mini \cite{openai_o3_mini} & 0.0171 & 0.1673 & 5.22\% & 0.5819 & -10.64\% & 0.0021 & 0.0242 & 3.31\% & 0.2404 & -19.94\% \\
    Deepseek-V3.1 \cite{deepseek_v3_1_release} & 0.0184 & 0.1758 & 4.93\% & 0.4861 & -16.71\% & 0.0026 & 0.0246 & 3.73\% & 0.2270 & -20.40\% \\
    AlphaForge \cite{shi2025alphaforge} & 0.0111 & 0.1345 & 3.15\% & 0.3020 & -25.28\% & 0.0026 & 0.0326 & 2.13\% & 0.3130 & -28.00\% \\
    RD-Agent \cite{li2025rd_agent_quant} & 0.0112 & 0.0966 & 1.01\% & 0.0930 & -22.27\% & 0.0019 & 0.0165 & 1.61\% & 0.1873 & -17.73\% \\
    AlphaAgent \cite{tang2025alphaagent} & 0.0221 & 0.2092 & 12.46\% & 1.2230 & -6.65\% & 0.0060 & 0.0515 & 8.57\% & 0.9653 & -9.44\% \\
    \textbf{AlphaLogics} & \textbf{0.0251} & \textbf{0.2312} & \textbf{16.72\%} & \textbf{1.5266} & \textbf{-5.31\%} & \textbf{0.0093} & \textbf{0.0878} & \textbf{13.75\%} & \textbf{1.2658} & \textbf{-9.06\%} \\
    \bottomrule
  \end{tabular}
\end{table*}

Table~\ref{tab:main_results} shows that AlphaLogics is best on IC/ICIR and AR/IR with the lowest MDD in both markets. In CSI 500, IR reaches 1.5266 versus 1.2230 for AlphaAgent; in S\&P 500, IR reaches 1.2658 and remains the best among baselines. The joint improvement on IC/ICIR and AR/IR indicates that AlphaLogics consistently delivers stronger rank-correlation signals and more stable portfolio outcomes than all baselines.

\subsection{Validity of Market Logic Mining: Reconstructing Factor Behavior from Explanations}
We assess extraction accuracy by reconstructing factor behavior from LLM explanations.

Following Section~\ref{subsec:logic_mining}, we extract market logic from historical factor libraries using a three-stage pipeline (mathematical formula, financial explanation, logic abstraction). Because market logic has no direct ground truth, we evaluate whether LLM explanations can reconstruct factor behavior.

We rebuild factor formulas from explanations and compare their outputs to the originals. Two formulas are equivalent if their cross-sectional rankings and time-series trends match in over 90\% of cases. We repeat this 100 times using Gemini-2.5-Flash and report consistency.

\begin{table}[!htbp]
  \caption{Consistency rates between original factor formulas and formulas reconstructed from LLM-provided interpretations. \textbf{Math. Expl.} denotes reconstruction consistency based on mathematical explanations of factor formulas (e.g., window length, smoothing, ranking, and normalization operations), while \textbf{Fin. Expl.} denotes consistency based on financial explanations mapping factors to economic meaning and market behavior.}
  \label{tab:hypothesis_accuracy}
  \centering
  \setlength{\tabcolsep}{6pt}
  \begin{tabular}{lcc}
    \toprule
    Factor Library & Math. Expl. & Fin. Expl. \\
    \midrule
    Alpha101  & 97.5\% & 92.7\% \\
    Alpha158  & 98.1\% & 95.5\% \\
    Alpha360  & 100.0\% & 98.8\% \\
    Alpha191  & 94.9\% & 93.8\% \\
    \bottomrule
  \end{tabular}
\end{table}

Table~\ref{tab:hypothesis_accuracy} shows consistency above 95\% for math explanations and 92--99\% for financial explanations, supporting accurate extraction of market logic from historical factor libraries. This accuracy is essential because market logic serves as the control signal for downstream factor generation. High reconstruction consistency suggests that the mining pipeline preserves both the structural and semantic aspects of the original factors, enabling the market logic library to provide faithful guidance in later optimization stages.

\subsection{Ablation: Effect of Executable Logic Constraints ($\Gamma$)}
We test whether structured market logic constraints improve factor quality.

We compare factor generation with and without executable market-logic constraints. In the constrained setting, we compile $H^{\text{struct}}$ into $\Gamma$ and enforce $\Gamma$ via DSL parsing and constraint checking; invalid programs are rejected and regenerated. In the unconstrained setting, we remove $\Gamma$ and allow free composition within the same DSL operator set and parameter ranges, keeping prompts and budgets otherwise identical. Each setting is repeated 20 times and averaged.

\begin{figure}[t]
  \centering
  \includegraphics[width=\columnwidth]{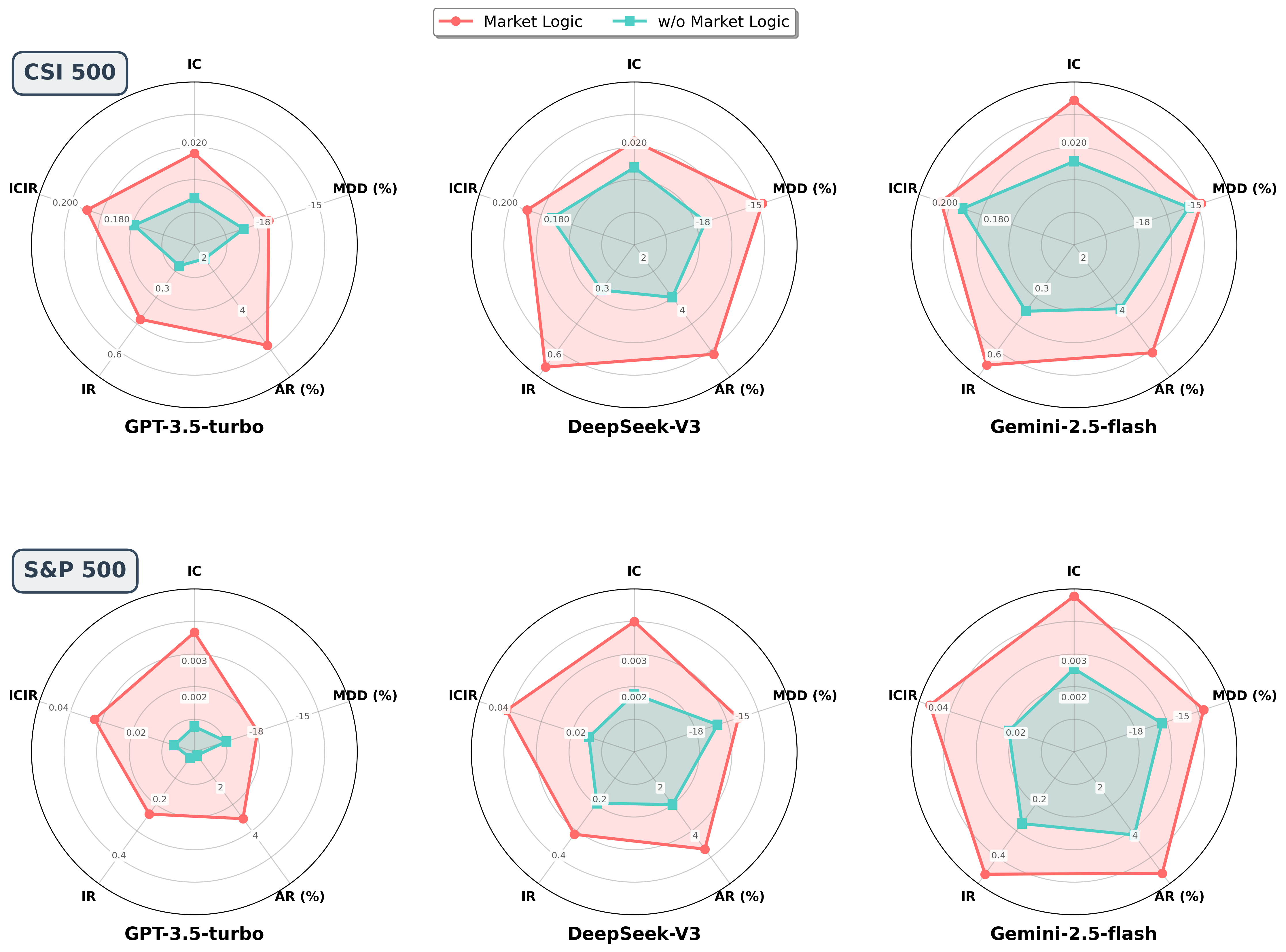}
  \caption{Comparison of market logic-guided factor generation on CSI 500 and S\&P 500 (train 2015.01--2019.12, validation 2020.01--2020.12, test 2021.01--2024.12). The results demonstrate that across all models and market settings, factors generated with market logic guidance consistently outperform unconstrained generation in terms of IC and IR metrics.}
  \label{fig:hypothesis_guided}
\end{figure}

We test CSI 500 and S\&P 500 with GPT-3.5-Turbo \cite{ouyang2022training}, DeepSeek V3 \cite{liu2024deepseek_v3}, and Gemini-2.5-Flash \cite{comanici2025gemini}. Figure~\ref{fig:hypothesis_guided} shows consistent gains in IC and IR across models and markets, indicating that market logic reduces unproductive exploration and improves stability.

The constrained setting filters the search space by enforcing logic-consistent operators and behavioral assumptions, while the unconstrained setting freely composes expressions without such structure. The gap between the two highlights that even strong LLMs benefit from explicit market logic, which helps prioritize economically meaningful patterns over spurious correlations.

\subsection{Analysis: Benefit of Outer-Loop Market Logic Refinement}
We test whether iterative logic optimization improves factor outcomes.

We evaluate the evolution of market logic using $H=\langle \mathcal{C}, \mathcal{B}\rangle$, where higher-quality logic should stably yield better factors under fixed constraints.

\begin{figure*}[t]
  \centering
  \includegraphics[width=0.88\textwidth]{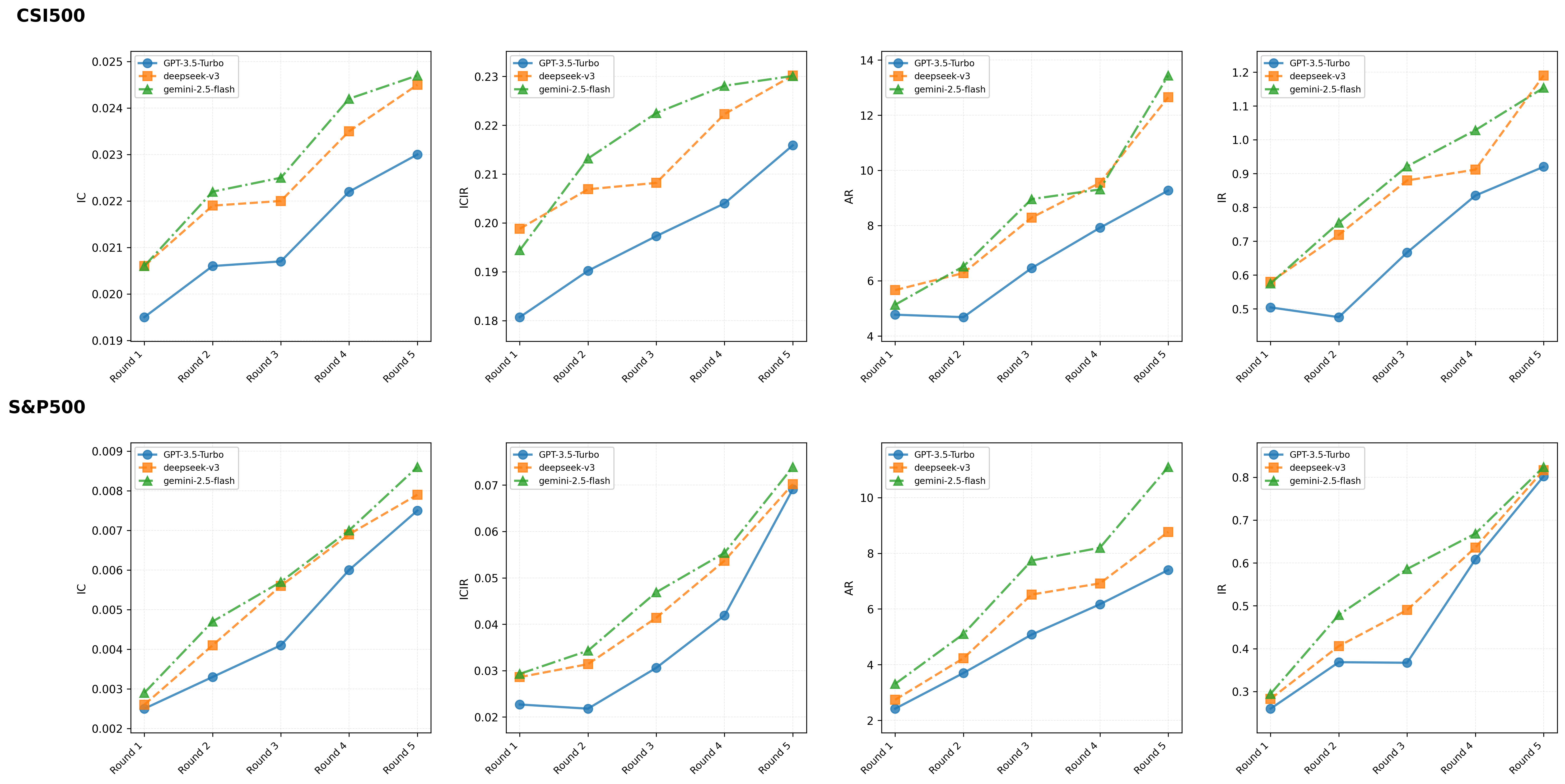}
  \caption{Evolution indicators of market logic. We test GPT-3.5-Turbo \cite{ouyang2022training}, DeepSeek V3 \cite{liu2024deepseek_v3}, and Gemini-2.5-Flash \cite{comanici2025gemini} on CSI 500 and S\&P 500, using early stopping of 3 for each logic and selecting the best-performing factor under the current logic in each round. The results show that as iteration rounds progress, the optimal factors corresponding to different rounds show an overall upward trend in key metrics such as IC, IR, cumulative returns, and stability.}
  \label{fig:evolution}
\end{figure*}

Figure~\ref{fig:evolution} shows upward trends in IC, IR, annual returns, and stability as rounds progress. Stronger LLMs help overall performance, while the optimization loop consistently improves market logic. The outer-loop refinement aggregates performance across all factors generated under the same market logic, producing refinement suggestion at the logic level rather than at the single-factor level. This encourages market logics refinement (e.g., tightening conditions or emphasizing specific behaviors) instead of merely searching for a better expression, enabling the system to improve the underlying market logic over time.

\subsection{Component Study: Scaling with Logic Library Size \& Cross-Round Persistence}
We ablate library size and cross-round persistence to test which components drive stability.

We vary the number of market logic instances from 1 to 6 under the same pipeline and backtesting setup, using GPT-5-mini \cite{openai_gpt5_mini}, to test whether a small logic set is sufficient.

\begin{figure}[!t]
  \centering
  \includegraphics[width=\columnwidth,trim=10 70 10 40,clip]{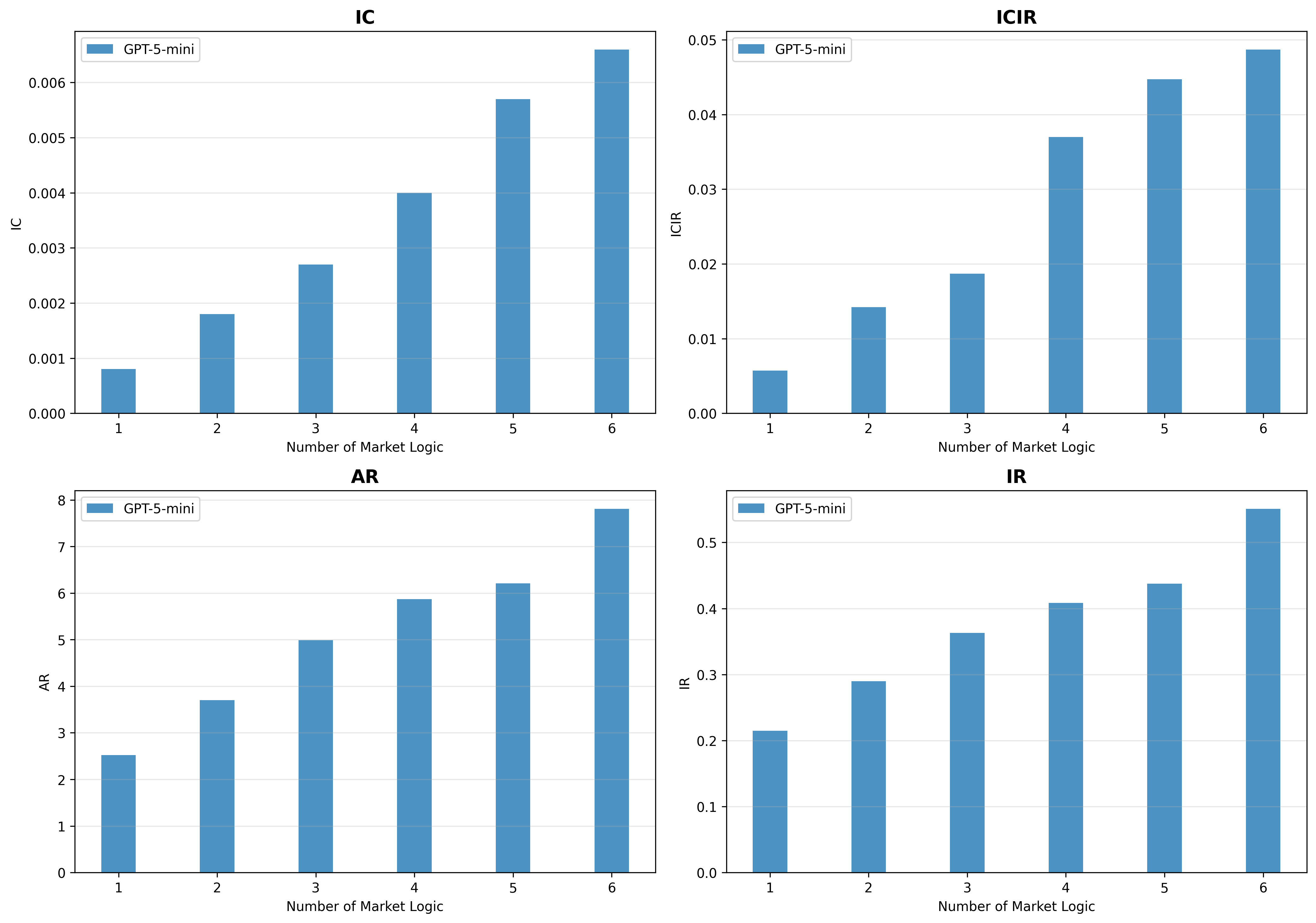}
  \caption{\small The influence of market logic quantity on factor quality. Increasing the number of market logic yields stable and nearly monotonic improvements across IC, ICIR, AR, and IR.}
  \label{fig:ablation}
\end{figure}

Figure~\ref{fig:ablation} shows stable, near-monotonic gains in IC, ICIR, AR, and IR as the library grows, supporting the need for a larger market logic library.

We compare \textit{transient market logic} (single-round usage) with \textit{persistent market logic} (reused and optimized across rounds) to test the necessity of the Inner Loop (Algorithm~\ref{alg:inner-loop}).

\begin{table}[!htbp]
  \caption{Ablation study on market logic persistence for verifying the necessity of the Inner Loop design (Algorithm~\ref{alg:inner-loop}). The \textit{transient} setting applies each market logic to guide factor generation only once, whereas the \textit{persistent} setting reuses the same market logic across multiple rounds to enable deeper exploration.}
  \label{tab:ablation}
  \centering
  \small
  \setlength{\tabcolsep}{3pt}
  \begin{tabular}{ccccc}
    \toprule
    \textbf{Round} & \multicolumn{2}{c}{\textbf{IC}} & \multicolumn{2}{c}{\textbf{ICIR}} \\
    \cmidrule(lr){2-3}\cmidrule(lr){4-5}
     & Transient & Persistent & Transient & Persistent \\
    \midrule
    1 & 0.0182 & 0.0199 & 0.1776 & 0.1792 \\
    2 & 0.0181 & 0.0208 & 0.1794 & 0.1951 \\
    3 & 0.0188 & 0.0214 & 0.1767 & 0.2033 \\
    4 & 0.0169 & 0.0222 & 0.1556 & 0.2071 \\
    5 & 0.0165 & 0.0232 & 0.1320 & 0.2137 \\
    \bottomrule
  \end{tabular}
\end{table}

Table~\ref{tab:ablation} shows that transient logic degrades over rounds, while persistent logic improves steadily. Persisting market logic and accumulating feedback are critical for fully exploring logic potential (Appendix~\ref{subsec:app_evolution}).

The library-size ablation further indicates that expanding the market logic inventory increases the diversity of market logic available for factor construction, reducing the chance that the search is bottlenecked by an overly narrow market logic set. Combined with persistence across rounds, AlphaLogics can both explore a broader space of candidate market logics and deepen evaluation for each market logic over time.

\section{Conclusion}

We propose AlphaLogics, a market logic-driven multi-agent system for scalable and interpretable alpha factor generation, integrating market logic extraction, feedback-based optimization, and logic-guided factor generation into a unified pipeline. Experiments across multiple historical factor libraries show that explicitly modeling and iteratively optimizing market logic improves both factor quality and stability, while enabling interpretable explanations of generated factors.

AlphaLogics provides an explicit market-logic layer that improves interpretability and auditability in automated factor research, enabling practitioners to reason about why a factor should work and how it generalizes across regimes. The framework can be applied to quantitative research pipelines and systematic factor discovery in portfolio construction, especially where transparent factor rationale is required. By treating market logic as a reusable asset, AlphaLogics also supports continual expansion of logic libraries for scalable alpha mining.

\section*{Impact Statement}

This work advances LLM-driven factor discovery in quantitative finance by modeling market logic as an interpretable and optimizable intermediate representation, which may improve transparency, auditability, and diagnostic analysis in automated factor research and portfolio modeling. Potential applications include quantitative research and systematic investment workflows; however, large-scale deployment of LLM-based automated factor generation and trading systems could influence market dynamics under widespread adoption of similar strategies. We do not advocate indiscriminate production use, and instead view explicit logic modeling as a step toward more responsible deployment by enabling practitioners to better understand, evaluate, and constrain LLM-driven decisions, while future work should further examine broader economic, regulatory, and systemic implications.

\begingroup
\raggedbottom
\bibliographystyle{icml2026}
\bibliography{references}

@article{gu2020empirical,
  title={Empirical asset pricing via machine learning},
  author={Gu, Shihao and Kelly, Bryan and Xiu, Dacheng},
  journal={The Review of Financial Studies},
  volume={33},
  number={5},
  pages={2223--2273},
  year={2020},
  publisher={Oxford University Press}
}

@article{zhang2020autoalpha,
  title={Autoalpha: an efficient hierarchical evolutionary algorithm for mining alpha factors in quantitative investment},
  author={Zhang, Tianping and Li, Yuanqi and Jin, Yifei and Li, Jian},
  journal={arXiv preprint arXiv:2002.08245},
  year={2020}
}

@article{fama2015five,
  title={A five-factor asset pricing model},
  author={Fama, Eugene F and French, Kenneth R},
  journal={Journal of financial economics},
  volume={116},
  number={1},
  pages={1--22},
  year={2015},
  publisher={Elsevier}
}

@misc{guotai191alphafactor,
  title={Guotai Junan 191 Alpha Factor Library: Technical Report and Implementation Code},
  author={{Guotai Junan Securities}},
  howpublished={GitHub Repository},
  note={\url{https://github.com/SelenaMa9812/Guotai-Junan-191-Alpha}},
  year={2025},
  key={Guotai-Junan-191-Alpha}
}

@inproceedings{cui2021alphaevolve,
  title={Alphaevolve: A learning framework to discover novel alphas in quantitative investment},
  author={Cui, Can and Wang, Wei and Zhang, Meihui and Chen, Gang and Luo, Zhaojing and Ooi, Beng Chin},
  booktitle={Proceedings of the 2021 International conference on management of data},
  pages={2208--2216},
  year={2021}
}

@article{kakushadze2016,
  title={101 formulaic alphas},
  author={Kakushadze, Zura},
  journal={Wilmott},
  volume={2016},
  number={84},
  pages={72--81},
  year={2016},
  publisher={Wiley Online Library}
}

@article{ren2024alphamining,
  title={Alpha Mining and Enhancing via Warm Start Genetic Programming for Quantitative Investment},
  author={Ren, Weizhe and Qin, Yichen and Li, Yang},
  journal={arXiv preprint arXiv:2412.00896},
  year={2024}
}

@inproceedings{shi2025alphaforge,
  title={Alphaforge: A framework to mine and dynamically combine formulaic alpha factors},
  author={Shi, Hao and Song, Weili and Zhang, Xinting and Shi, Jiahe and Luo, Cuicui and Ao, Xiang and Arian, Hamid and Seco, Luis Angel},
  booktitle={Proceedings of the AAAI Conference on Artificial Intelligence},
  volume={39},
  number={12},
  pages={12524--12532},
  year={2025}
}

@article{tatsat2025beyond,
  title={Beyond the black box: Interpretability of llms in finance},
  author={Tatsat, Hariom and Shater, Ariye},
  journal={arXiv preprint arXiv:2505.24650},
  year={2025}
}

@article{rudin2019why,
  title={Why are we using black box models in AI when we don’t need to? A lesson from an explainable AI competition},
  author={Rudin, Cynthia and Radin, Joanna},
  journal={Harvard Data Science Review},
  volume={1},
  number={2},
  pages={1--9},
  year={2019},
  publisher={The MIT Press}
}

@article{tong2024ploutos,
  title={Ploutos: Towards interpretable stock movement prediction with financial large language model},
  author={Tong, Hanshuang and Li, Jun and Wu, Ning and Gong, Ming and Zhang, Dongmei and Zhang, Qi},
  journal={arXiv preprint arXiv:2403.00782},
  year={2024}
}

@inproceedings{wang2025alpha_gpt,
  title={Alpha-gpt: Human-ai interactive alpha mining for quantitative investment},
  author={Wang, Saizhuo and Yuan, Hang and Zhou, Leon and Ni, Lionel and Shum, Heung Yeung and Guo, Jian},
  booktitle={Proceedings of the 2025 Conference on Empirical Methods in Natural Language Processing: System Demonstrations},
  pages={196--206},
  year={2025}
}

@inproceedings{duan2025factormad,
  title={FactorMAD: A Multi-Agent Debate Framework Based on Large Language Models for Interpretable Stock Alpha Factor Mining},
  author={Duan, Yitong and zhang, chuheng and Li, Jian},
  booktitle={Proceedings of the 6th ACM International Conference on AI in Finance},
  pages={605--613},
  year={2025}
}

@article{nie2024survey_llm_finance,
  title={A survey of large language models for financial applications: Progress, prospects and challenges},
  author={Nie, Yuqi and Kong, Yaxuan and Dong, Xiaowen and Mulvey, John M and Poor, H Vincent and Wen, Qingsong and Zohren, Stefan},
  journal={arXiv preprint arXiv:2406.11903},
  year={2024}
}

@article{ding2023integrating,
  title={Integrating stock features and global information via large language models for enhanced stock return prediction},
  author={Ding, Yujie and Jia, Shuai and Ma, Tianyi and Mao, Bingcheng and Zhou, Xiuze and Li, Liuliu and Han, Dongming},
  journal={arXiv preprint arXiv:2310.05627},
  year={2023}
}

@article{afzal2023volatility,
  title={Volatility spillover effect between Pakistan and Shanghai Stock Exchanges using copula and dynamic conditional correlation model},
  author={Afzal, Fahim and Choudhury, Tonmoy Toufic and Kamran, Muhammad},
  journal={International Journal of Islamic and Middle Eastern Finance and Management},
  volume={16},
  number={1},
  pages={59--80},
  year={2023},
  publisher={Emerald Publishing Limited}
}

@misc{hibbeln2025model_validation,
  title={Model Validation for Forecasting: Out-of-sample vs. Out-of-time},
  author={Hibbeln, Martin Thomas and Jentsch, Carsten and Kopp, Raphael M and Urban, Noah},
  journal={Out-of-time (April 29, 2025)},
  year={2025}
}

@article{fama1993common,
  title={Common risk factors in the returns on stocks and bonds},
  author={Fama, Eugene F and French, Kenneth R},
  journal={Journal of financial economics},
  volume={33},
  number={1},
  pages={3--56},
  year={1993},
  publisher={Elsevier}
}

@article{fama2018choosing,
  title={Choosing factors},
  author={Fama, Eugene F and French, Kenneth R},
  journal={Journal of financial economics},
  volume={128},
  number={2},
  pages={234--252},
  year={2018},
  publisher={Elsevier}
}

@article{ye2024from_factor_models,
  title={From factor models to deep learning: Machine learning in reshaping empirical asset pricing},
  author={Ye, Junyi and Goswami, Bhaskar and Gu, Jingyi and Uddin, Ajim and Wang, Guiling},
  journal={arXiv preprint arXiv:2403.06779},
  year={2024}
}

@article{liu2025deep_conditional,
  title={Deep Learning for Conditional Asset Pricing Models},
  author={Liu, Hongyi},
  journal={arXiv preprint arXiv:2509.04812},
  year={2025}
}

@article{feng2024deep_learning_character,
  title={Deep learning in characteristics-sorted factor models},
  author={Feng, Guanhao and He, Jingyu and Polson, Nicholas G and Xu, Jianeng},
  journal={Journal of Financial and Quantitative Analysis},
  volume={59},
  number={7},
  pages={3001--3036},
  year={2024},
  publisher={Cambridge University Press}
}

@article{liao2025uncertainty_ml,
  title={The Uncertainty of Machine Learning Predictions in Asset Pricing},
  author={Liao, Yuan and Ma, Xinjie and Neuhierl, Andreas and Schilling, Linda},
  journal={arXiv preprint arXiv:2503.00549},
  year={2025}
}

@article{kou2024automate_strategy,
  title={Automate strategy finding with llm in quant investment},
  author={Kou, Zhizhuo and Yu, Holam and Luo, Junyu and Peng, Jingshu and Li, Xujia and Liu, Chengzhong and Dai, Juntao and Chen, Lei and Han, Sirui and Guo, Yike},
  journal={arXiv preprint arXiv:2409.06289},
  year={2024}
}

@inproceedings{tang2025alphaagent,
  title={AlphaAgent: LLM-driven alpha mining with regularized exploration to counteract alpha decay},
  author={Tang, Ziyi and Chen, Zechuan and Yang, Jiarui and Mai, Jiayao and Zheng, Yongsen and Wang, Keze and Chen, Jinrui and Lin, Liang},
  booktitle={Proceedings of the 31st ACM SIGKDD Conference on Knowledge Discovery and Data Mining V. 2},
  pages={2813--2822},
  year={2025}
}

@article{yang2020qlib,
  title={Qlib: An ai-oriented quantitative investment platform},
  author={Yang, Xiao and Liu, Weiqing and Zhou, Dong and Bian, Jiang and Liu, Tie-Yan},
  journal={arXiv preprint arXiv:2009.11189},
  year={2020}
}

@misc{baostock2024,
  title={BaoStock: A Tool for Obtaining Historical Data of China Stock Market},
  author={{BaoStock}},
  howpublished={\url{https://pypi.org/project/baostock/}},
  year={2024}
}

@misc{ranaroussi2024yfinance,
  title={yfinance: Download Market Data from Yahoo! Finance's API},
  author={Ran Aroussi},
  howpublished={\url{https://pypi.org/project/yfinance/}},
  year={2024}
}

@incollection{graves2012lstm,
  title={Long short-term memory},
  author={Graves, Alex},
  journal={Supervised sequence labelling with recurrent neural networks},
  pages={37--45},
  year={2012},
  publisher={Springer}
}

@article{chung2014gru,
  title={Empirical evaluation of gated recurrent neural networks on sequence modeling},
  author={Chung, Junyoung and Gulcehre, Caglar and Cho, KyungHyun and Bengio, Yoshua},
  journal={arXiv preprint arXiv:1412.3555},
  year={2014}
}

@incollection{taud2017mlp,
  title={Multilayer perceptron (MLP)},
  author={Taud, Hind and Mas, Jean-Franccois},
  booktitle={Geomatic approaches for modeling land change scenarios},
  pages={451--455},
  year={2017},
  publisher={Springer}
}

@inproceedings{ke2017lightgbm,
  title={Lightgbm: A highly efficient gradient boosting decision tree},
  author={Ke, Guolin and Meng, Qi and Finley, Thomas and Wang, Taifeng and Chen, Wei and Ma, Weidong and Ye, Qiwei and Liu, Tie-Yan},
  journal={Advances in neural information processing systems},
  volume={30},
  year={2017}
}

@article{chen2016xgboost,
  title={XGBoost: A Scalable Tree Boosting System},
  author={Chen, Tianqi},
  journal={Cornell University},
  year={2016}
}

@inproceedings{vaswani2017attention,
  title={Attention is all you need},
  author={Vaswani, Ashish and Shazeer, Noam and Parmar, Niki and Uszkoreit, Jakob and Jones, Llion and Gomez, Aidan N and Kaiser, {\L}ukasz and Polosukhin, Illia},
  journal={Advances in neural information processing systems},
  volume={30},
  year={2017}
}

@inproceedings{lin2021temporal,
  title={Learning multiple stock trading patterns with temporal routing adaptor and optimal transport},
  author={Lin, Hengxu and Zhou, Dong and Liu, Weiqing and Bian, Jiang},
  booktitle={Proceedings of the 27th ACM SIGKDD conference on knowledge discovery \& data mining},
  pages={1017--1026},
  year={2021}
}

@article{li2025rd_agent_quant,
  title={R\&D-Agent-Quant: A Multi-Agent Framework for Data-Centric Factors and Model Joint Optimization},
  author={Li, Yuante and Yang, Xu and Yang, Xiao and Xu, Minrui and Wang, Xisen and Liu, Weiqing and Bian, Jiang},
  journal={arXiv preprint arXiv:2505.15155},
  year={2025}
}

@article{comanici2025gemini,
  title={Gemini 2.5: Advanced Reasoning, Multimodality, and Long Context Capabilities},
  author={Comanici, G. and Bieber, E. and Schaekermann, M. and others},
  journal={arXiv preprint arXiv:2507.06261},
  year={2025}
}

@article{ouyang2022training,
  title={Training language models to follow instructions with human feedback},
  author={Ouyang, Long and Wu, Jeffrey and Jiang, Xu and Almeida, Diogo and Wainwright, Carroll and Mishkin, Pamela and Zhang, Chong and Agarwal, Sandhini and Slama, Katarina and Ray, Alex and others},
  journal={Advances in neural information processing systems},
  volume={35},
  pages={27730--27744},
  year={2022}
}

@article{liu2024deepseek_v3,
  title={DeepSeek-v3 Technical Report},
  author={Liu, A. and Feng, B. and Xue, B. and others},
  journal={arXiv preprint arXiv:2412.19437},
  year={2024}
}

@misc{qlib_alpha158,
  title={Alpha158 Factor Library: Data Handler in Microsoft Qlib},
  author={{Qlib Team}},
  howpublished={Qlib Documentation, \url{https://qlib.readthedocs.io/en/latest/component/data.html}},
  year={2025}
}

@misc{qlib_alpha360,
  title={Alpha360 Factor Library: High-Dimensional Factor Dataset in Microsoft Qlib},
  author={{Qlib Team}},
  howpublished={Qlib Documentation, \url{https://qlib.readthedocs.io/en/latest/component/data.html}},
  year={2025}
}

@misc{openai_o3_mini,
  title={o3-mini: Efficient Reasoning Model for Fast and Cost-Effective Inference},
  author={{OpenAI}},
  howpublished={\url{https://platform.openai.com/docs/models/o3-mini}},
  year={2025}
}

@misc{deepseek_v3_1_release,
  title        = {DeepSeek-V3.1 Release},
  author       = {{DeepSeek}},
  howpublished = {DeepSeek API Docs (News)},
  note         = {\url{https://api-docs.deepseek.com/news/news250821} (accessed 2026-01-19)},
  year         = {2025}
}

@misc{openai_gpt5_mini,
  title        = {GPT-5 mini Model},
  author       = {{OpenAI}},
  howpublished = {OpenAI API Documentation (Models)},
  note         = {\url{https://platform.openai.com/docs/models/gpt-5-mini} (accessed 2026-01-19)},
  year         = {2025}
}
\endgroup

\newpage
\appendix
\onecolumn
\section{Appendix}

\subsection{Public Factor Libraries}

To validate the effectiveness of the proposed factor interpretation–market-logic evolution–factor generation framework, we select four widely recognized public factor libraries from both academia and industry as the empirical foundation of our study. These libraries cover technical factors with varying levels of complexity and distinct design philosophies, enabling a comprehensive evaluation of the framework across market logic extraction, market logic evolution, and new factor generation.

\begin{table}[!htbp]
  \caption{Overview of Major Public Factor Libraries.}
  \label{tab:factor_libraries}
  \centering
  \small
  \setlength{\tabcolsep}{4pt}
  \begin{tabularx}{\columnwidth}{l l c X}
    \toprule
    \textbf{Factor Library} & \textbf{Institution} & \textbf{number} & \textbf{Core Characteristics} \\
    \midrule
    Alpha101 & WorldQuant & 101 &
    Uses ranking, lag, and cross-sectional normalization; focuses on momentum, reversal, and price--volume interaction. \\
    Alpha191 & Guotai Junan & 191 &
    Covers momentum, volatility, microstructure, and trading behavior. \\
    Alpha158 & Microsoft Research & 158 &
    Rich operator set with strong cross-market generalization. \\
    Alpha360 & Microsoft Research & 360 &
    Largest public technical factor library; encompasses price, volume, volatility, and technical indicators. \\
    \bottomrule
  \end{tabularx}
\end{table}

The four types of factor libraries play two core roles in this study: Firstly, as the input source for the deconstruction of the ``factor-market logic'' relationship; Secondly, as the initial population for the evolution of market logic.

\subsection{Factor DSL and Parameter Ranges}
\label{subsec:app_factor_dsl}

Factor expressions are generated by composing operators from a fixed DSL. We group operators into families and treat their parameters as positive integers without global bounds; in practice they are limited only by data availability.

\begin{table}[!htbp]
  \caption{Operator families and parameter ranges used in the factor DSL.}
  \label{tab:factor_dsl}
  \centering
  \small
  \setlength{\tabcolsep}{5pt}
  \begin{tabular}{l l l}
    \toprule
    \textbf{Family} & \textbf{Examples} & \textbf{Parameters (range)} \\
    \midrule
    Arithmetic & $+$, $-$, $\times$, $/$ & -- \\
    Cross-sectional & rank, zscore & -- \\
    Time-series aggregation & ts\_mean, ts\_std, ts\_min, ts\_max, ts\_sum & window $w \in \mathbb{N}^+$ \\
    Time-series change & ts\_delta & lag $\ell \in \mathbb{N}^+$ \\
    Time-series relation & ts\_corr, ts\_cov & window $w \in \mathbb{N}^+$ \\
    Smoothing/decay & ts\_decay, ts\_wma & window $w \in \mathbb{N}^+$ \\
    \bottomrule
  \end{tabular}
\end{table}

\subsection{Factor Operations Library}
\label{app:factor_ops}

We provide the full set of operations allowed in factor expressions. These concrete operators instantiate the families in Table~\ref{tab:factor_dsl}; expressions may only use the operations below.

\textbf{Cross-sectional functions.}
\begin{itemize}
  \item \texttt{RANK(A)}: Ranking of each element in the cross-sectional dimension of A.
  \item \texttt{ZSCORE(A)}: Z-score of each element in the cross-sectional dimension of A.
  \item \texttt{MEAN(A)}: Mean value of each element in the cross-sectional dimension of A.
  \item \texttt{STD(A)}: Standard deviation in the cross-sectional dimension of A.
  \item \texttt{SKEW(A)}: Skewness in the cross-sectional dimension of A.
  \item \texttt{KURT(A)}: Kurtosis in the cross-sectional dimension of A.
  \item \texttt{MAX(A)}: Maximum value in the cross-sectional dimension of A.
  \item \texttt{MIN(A)}: Minimum value in the cross-sectional dimension of A.
  \item \texttt{MEDIAN(A)}: Median value in the cross-sectional dimension of A.
\end{itemize}

\textbf{Time-series functions.}
\begin{itemize}
  \item \texttt{DELTA(A, n)}: Change in value of A over $n$ periods.
  \item \texttt{DELAY(A, n)}: Value of A delayed by $n$ periods.
  \item \texttt{TS\_MEAN(A, n)}: Mean value of sequence A over the past $n$ days.
  \item \texttt{TS\_SUM(A, n)}: Sum of sequence A over the past $n$ days.
  \item \texttt{TS\_RANK(A, n)}: Time-series rank of the last value of A in the past $n$ days.
  \item \texttt{TS\_ZSCORE(A, n)}: Z-score for each sequence in A over the past $n$ days.
  \item \texttt{TS\_MEDIAN(A, n)}: Median value of sequence A over the past $n$ days.
  \item \texttt{TS\_PCTCHANGE(A, p)}: Percentage change in the value of sequence A over $p$ periods.
  \item \texttt{TS\_MIN(A, n)}: Minimum value of A in the past $n$ days.
  \item \texttt{TS\_MAX(A, n)}: Maximum value of A in the past $n$ days.
  \item \texttt{TS\_ARGMAX(A, n)}: Index (relative to the current time) of the maximum value of A over the past $n$ days.
  \item \texttt{TS\_ARGMIN(A, n)}: Index (relative to the current time) of the minimum value of A over the past $n$ days.
  \item \texttt{TS\_QUANTILE(A, p, q)}: Rolling quantile of sequence A over the past $p$ periods, where $q \in (0, 1)$.
  \item \texttt{TS\_STD(A, n)}: Standard deviation of sequence A over the past $n$ days.
  \item \texttt{TS\_VAR(A, p)}: Rolling variance of sequence A over the past $p$ periods.
  \item \texttt{TS\_COVARIANCE(A, B, n)}: Covariance between sequences A and B over the past $n$ days.
  \item \texttt{TS\_MAD(A, n)}: Rolling median absolute deviation of sequence A over the past $n$ days.
  \item \texttt{PERCENTILE(A, q, p)}: Quantile of sequence A with $q \in (0, 1)$; if $p$ is provided, compute the rolling quantile over the past $p$ periods.
  \item \texttt{HIGHDAY(A, n)}: Number of days since the highest value of A in the past $n$ days.
  \item \texttt{LOWDAY(A, n)}: Number of days since the lowest value of A in the past $n$ days.
  \item \texttt{SUMAC(A, n)}: Cumulative sum of A over the past $n$ days.
\end{itemize}

\textbf{Moving averages and smoothing functions.}
\begin{itemize}
  \item \texttt{SMA(A, n, m)}: Simple moving average of A over $n$ periods with modifier $m$.
  \item \texttt{WMA(A, n)}: Weighted moving average of A over $n$ periods, with weights decreasing from $0.9$ to $0.9^{n}$.
  \item \texttt{EMA(A, n)}: Exponential moving average of A over $n$ periods, with decay factor $2/(n+1)$.
  \item \texttt{DECAYLINEAR(A, d)}: Linearly weighted moving average of A over $d$ periods, with weights increasing from $1$ to $d$.
\end{itemize}

\textbf{Mathematical operations.}
\begin{itemize}
  \item \texttt{PROD(A, n)}: Product of values in A over the past $n$ days; use \texttt{*} for general multiplication.
  \item \texttt{LOG(A)}: Natural logarithm of each element in A.
  \item \texttt{SQRT(A)}: Square root of each element in A.
  \item \texttt{POW(A, n)}: Raise each element in A to the power of $n$.
  \item \texttt{SIGN(A)}: Sign of each element in A (one of 1, 0, or -1).
  \item \texttt{EXP(A)}: Exponential of each element in A.
  \item \texttt{ABS(A)}: Absolute value of A.
  \item \texttt{MAX(A, B)}: Maximum value between A and B.
  \item \texttt{MIN(A, B)}: Minimum value between A and B.
  \item \texttt{INV(A)}: Reciprocal ($1/x$) of each element in sequence A.
  \item \texttt{FLOOR(A)}: Floor of each element in sequence A.
\end{itemize}

\textbf{Conditional and logical functions.}
\begin{itemize}
  \item \texttt{COUNT(C, n)}: Count of samples satisfying condition C in the past $n$ periods.
  \item \texttt{SUMIF(A, n, C)}: Sum of A over the past $n$ periods if condition C is met.
  \item \texttt{FILTER(A, C)}: Filter multi-column sequence A based on condition C, with the same size as A.
  \item \texttt{(C1)\&\&(C2)}: Logical \texttt{and} between conditions C1 and C2 (e.g., \texttt{close > open}).
  \item \texttt{(C1)||(C2)}: Logical \texttt{or} between conditions C1 and C2.
  \item \texttt{(C1)?(A):(B)}: If condition C1 holds, return A; otherwise return B.
\end{itemize}

\textbf{Regression and residual functions.}
\begin{itemize}
  \item \texttt{SEQUENCE(n)}: Single-column sequence of length $n$ ranging from 1 to $n$; it should be nested in \texttt{REGBETA} or \texttt{REGRESI} as argument B.
  \item \texttt{REGBETA(A, B, n)}: Regression coefficient of A on B using the past $n$ samples, where A must be multi-column and B may be single- or multi-column.
  \item \texttt{REGRESI(A, B, n)}: Residual of regression of A on B using the past $n$ samples, where A must be multi-column and B may be single- or multi-column.
\end{itemize}

\textbf{Technical indicators.}
\begin{itemize}
  \item \texttt{RSI(A, n)}: Relative Strength Index of sequence A over $n$ periods.
  \item \texttt{MACD(A, short\_window, long\_window)}: Moving Average Convergence Divergence of sequence A, defined as the difference between short- and long-window EMAs.
  \item \texttt{BB\_MIDDLE(A, n)}: Middle Bollinger Band, the $n$-period SMA of sequence A.
  \item \texttt{BB\_UPPER(A, n)}: Upper Bollinger Band, middle band plus two standard deviations over $n$ periods.
  \item \texttt{BB\_LOWER(A, n)}: Lower Bollinger Band, middle band minus two standard deviations over $n$ periods.
\end{itemize}

\subsection{Logic Schema and Compilation Example}
\label{app:compile}

\textbf{Structured schema.}
We store $H^{\text{struct}}$ as a typed record with fields: (i) $C$, a Boolean formula over predicates $(v, \mathrm{op}, \theta, w)$; (ii) $B=(y,d,h)$ with target $y$, direction $d \in \{+1,-1\}$, and horizon $h$.

\textbf{Deterministic compilation and enforcement.}
$\Gamma = \textsc{Compile}(H^{\text{struct}})$ is a deterministic rule-based mapping from predicate types to allowed variable families, operator families in the factor DSL, parameter constraints, and sign/direction constraints. Generated programs are parsed by a DSL validator and checked against $\Gamma$; invalid programs are rejected and regenerated.

\textbf{Example.}
\noindent\textbf{Logic $H^{\text{struct}}$:}
\begin{verbatim}
C: (price_trend_up over w=1) AND (volume_trend_not_up over w=1)
B: (y = forward_return, d = -1, h = 1)
\end{verbatim}

\noindent\textbf{Compiled constraints $\Gamma$:}
\begin{itemize}
  \item Variables must include \{price, volume\}.
  \item Operator families: allow \{rank, zscore, ts\_delta, ts\_corr, ts\_mean\}.
  \item Parameter ranges: $w \in \mathbb{N}^+$, $\ell \in \mathbb{N}^+$ (Appendix~\ref{subsec:app_factor_dsl}).
  \item Direction: prefer factors with negative IC on $D_{\text{val}}$.
\end{itemize}

\subsection{Effectiveness Evaluation Metrics}
\label{subsec:app_metrics}

In the market logic - factor evolution framework, we measure the effectiveness of market logic by using the factors generated under its guidance. The evaluation of the factors needs to be conducted from three dimensions - ``predictive ability, return performance, and risk control'' through multiple indicators:

\textbf{Information Coefficient (IC):} The IC is computed using the Spearman rank correlation:
\begin{equation}
\text{IC}_t = \text{corr}_{\text{Spearman}}(f_t, r_{t+1})
\end{equation}
where $t$ indexes trading days, $f_t$ is the cross-sectional vector of factor scores for all tradable stocks at time $t$, $r_{t+1}$ is the corresponding vector of realized next-period returns, and $\text{corr}_{\text{Spearman}}(\cdot,\cdot)$ is computed across the cross section at each $t$. The final IC is obtained as the time-series average of $\text{IC}_t$.

\textbf{Information Coefficient Information Ratio (ICIR):} Defined as the ratio of the mean to the standard deviation of the IC time series:
\begin{equation}
\text{ICIR} = \frac{\mathbb{E}[\text{IC}]}{\sigma[\text{IC}]}
\end{equation}
where $\mathbb{E}[\text{IC}]$ is the time-series mean of $\text{IC}_t$ over the evaluation period, and $\sigma[\text{IC}]$ is the time-series standard deviation of $\text{IC}_t$.

\textbf{Annualized Return (AR):} Based on a factor-constructed investment portfolio:
\begin{equation}
\text{AR} = \bar{r} \times \text{annualization factor}
\end{equation}
where $r_t$ denotes the per-period portfolio (or excess) return at time $t$, $\bar{r} = \mathbb{E}[r_t]$ is its time-series mean, and the annualization factor is the number of trading periods per year (e.g., 252 for daily data).

\textbf{Information Ratio (IR):} Defined as the ratio of annualized excess return to annualized tracking error:
\begin{equation}
\text{IR} = \frac{\text{AR}}{\sigma[r] \times \sqrt{\text{annualization factor}}}
\end{equation}
where $\sigma[r]$ is the time-series standard deviation of per-period (excess) returns $r_t$, and $\sqrt{\text{annualization factor}}$ annualizes the tracking error; $\text{AR}$ is defined above.

\textbf{Maximum Drawdown (MDD):} The maximum drawdown during the backtesting period:
\begin{equation}
\text{MDD} = \max_{t} \left( \frac{\text{peak}_t - \text{trough}_t}{\text{peak}_t} \right)
\end{equation}
where $\text{peak}_t$ is the running maximum of the cumulative portfolio value up to time $t$, $\text{trough}_t$ is the minimum cumulative value observed after that peak within the drawdown episode, and the maximum is taken over the full backtest period.

\subsection{Market Logic Optimization Example}
\label{subsec:app_evolution}

To further illustrate the evolution process of persistent market logic, we present a representative logic trajectory optimized under our proposed framework (based on price--volume divergence and short-term reversal logic):

\begin{itemize}
  \item \textbf{Initial market logic:} ``(C) When an asset's intraday high increases but volume does not correspondingly rise; (B) the asset may experience a short-term price reversal in the next trading period.''
  \item \textbf{Iteration 1:} Added volume-to-K-line body ratio information: ``(C) When the closing price rises with declining volume, and the K-line body (closing price minus opening price) occupies a small proportion of daily price fluctuation; (B) this indicates insufficient upward momentum and a higher probability of reversal.''
  \item \textbf{Iteration 2:} Introduced gap signals: ``(C) When an asset gaps up (opening price higher than previous high) but volume shrinks, and the gap is not held (closing price below opening); (B) this `gap-up with volume shrink + bearish close' combination signals a strong reversal.''
  \item \textbf{Iteration 3:} Combined high-price and low-volume information: ``(C) When recent prices are at relatively high levels while volume is low; (B) this `high-price low-volume' state indicates weak buying support and signals a strong reversal.''
  \item \textbf{Iteration 4:} Considered divergence strength and timing of volume surges: ``(C) The first occurrence of volume contraction with price increase after a strong uptrend (divergence) serves as a reversal warning, and cumulative volume-price divergence over consecutive days strengthens the reversal signal; (B) the first bearish candle with increased volume post-divergence marks the optimal entry point.''
\end{itemize}

This optimization path demonstrates the ability of persistent market logic to gradually absorb factor feedback, refine signals, and combine conditions across iterations, highlighting the framework's explicit modeling and cross-round optimization of market logic.

\subsection{Agent Prompt Templates}
\label{subsec:app_prompts}

We provide JSON prompt templates and explicit output schemas for all agents in the pipeline.

\textbf{FormulaStructureAgent.}
\begin{tcblisting}{listing only, breakable, colback=white, colframe=black, boxrule=0.5pt, listing options={basicstyle=\ttfamily\footnotesize, breaklines=true}}
{
  "system": "Decompose the factor formula into operational structure and formal properties only. Identify sub-expressions, map operators to mathematical meaning using the factor operations library, and note any inferred operators. Use only variables {open, close, high, low, volume, return} with canonical meanings. Do not provide financial interpretation.",
  "instruction": "Identify sub-expressions, map operators to mathematical meaning using the factor operations library, and return a structured decomposition.",
  "input_schema": {
    "formula": "<string>",
    "factor_operations_library": "<string>"
  },
  "output_schema": {
    "components": [
      {
        "name": "<string>",
        "expression": "<string>",
        "mathematical_meaning": "<string>"
      }
    ]
  }
}
\end{tcblisting}

\textbf{FinancialSemanticsMappingAgent.}
\begin{tcblisting}{listing only, breakable, colback=white, colframe=black, boxrule=0.5pt, listing options={basicstyle=\ttfamily\footnotesize, breaklines=true}}
{
  "system": "Translate mathematical factor components into financial interpretations. Preserve mathematical_meaning exactly, and connect each component to market behavior, investor psychology, trading patterns, liquidity, information flow, and risk-return characteristics using the factor operations library.",
  "instruction": "Given factor_formula, mathematical_analysis, and factor_operations_library, add financial_interpretation to each component; preserve mathematical_meaning exactly and return only JSON. Use only variables {open, close, high, low, volume, return} with their standard meanings.",
  "input_schema": {
    "factor_formula": "<string>",
    "mathematical_analysis": {
      "components": [
        {
          "name": "<string>",
          "expression": "<string>",
          "mathematical_meaning": "<string>"
        }
      ]
    },
    "factor_operations_library": "<string>"
  },
  "output_schema": {
    "components": [
      {
        "name": "<string>",
        "expression": "<string>",
        "mathematical_meaning": "<string>",
        "financial_interpretation": "<string>"
      }
    ]
  }
}
\end{tcblisting}

\textbf{MarketLogicAbstractionAgent.}
\begin{tcblisting}{listing only, breakable, colback=white, colframe=black, boxrule=0.5pt, listing options={basicstyle=\ttfamily\footnotesize, breaklines=true}}
{
  "system": "Abstract component-level financial semantics into explicit market logic H with C/B semantics. Provide a concise, human-readable logic_text and explicit C (conditions) and B (target, direction, horizon) to support downstream canonicalization.",
  "instruction": "Given component_analysis, infer the market logic H. Provide a human-readable logic_text plus explicit C (conditions) and B (target/direction/horizon). Avoid formula details and keep the logic generalizable.",
  "input_schema": {
    "component_analysis": {
      "components": [
        {
          "name": "<string>",
          "expression": "<string>",
          "mathematical_meaning": "<string>",
          "financial_interpretation": "<string>"
        }
      ]
    }
  },
  "output_schema": {
    "logic_text": "<string>",
    "c_text": "<string>",
    "b_text": "<string>"
  }
}
\end{tcblisting}

\textbf{LogicToFinanceConstraintAgent.}
\begin{tcblisting}{listing only, breakable, colback=white, colframe=black, boxrule=0.5pt, listing options={basicstyle=\ttfamily\footnotesize, breaklines=true}}
{
  "system": "Canonicalize H into H_struct (C as a Boolean formula over predicates; B with target, direction, horizon), then compile constraints Gamma over variables, operator families, parameter ranges, and direction consistency.",
  "instruction": "Canonicalize H into H_struct (C as a Boolean formula over predicates, B with target/direction/horizon), then compile constraints Gamma.",
  "input_schema": {
    "logic_text": "<string>",
    "c_text": "<string>",
    "b_text": "<string>",
    "dsl_operators": ["<string>"]
  },
  "output_schema": {
    "H_struct": {
      "C": {
        "formula": "<string>",
        "predicates": [
          {"id": "<string>", "v": "<string>", "op": "<string>", "theta": "<string>", "w": "<int>"}
        ]
      },
      "B": {"y": "<string>", "d": "<+1|-1>", "h": "<int>"}
    },
    "Gamma": {
      "allowed_variables": ["<string>"],
      "operator_families": ["<string>"],
      "parameter_constraints": {"window": "positive integer", "lag": "positive integer"},
      "direction_constraint": "<string>"
    },
    "canonicalization_notes": "<string>"
  }
}
\end{tcblisting}

\textbf{FactorExpressionGeneratorAgent.}
\begin{tcblisting}{listing only, breakable, colback=white, colframe=black, boxrule=0.5pt, listing options={basicstyle=\ttfamily\footnotesize, breaklines=true}}
{
  "system": "Generate candidate factor expressions that satisfy Gamma. Select operators, time windows, and compositions consistent with allowed variables, operator families, and parameter constraints; ensure direction consistency.",
  "instruction": "Generate candidate factor expressions consistent with Gamma. Use allowed variables, operator families, parameter constraints, and direction constraints. Incorporate feedback if provided, and return up to max_candidates expressions with brief rationale and operator list.",
  "input_schema": {
    "Gamma": "<object>",
    "feedback": "<string or object or null>",
    "max_candidates": "<int>"
  },
  "output_schema": {
    "factors": [
      {"expression": "<string>", "rationale": "<string>", "operators": ["<string>"]}
    ],
    "notes": "<string>"
  }
}
\end{tcblisting}

\textbf{FactorPerformanceFeedbackAgent.}
\begin{tcblisting}{listing only, breakable, colback=white, colframe=black, boxrule=0.5pt, listing options={basicstyle=\ttfamily\footnotesize, breaklines=true}}
{
  "system": "Summarize candidate performance under a fixed logic. Compare validation metrics across candidates, identify the best expression, diagnose weaknesses, and suggest edits to guide the next generation.",
  "instruction": "Compare recent candidates under the same logic and validation metrics. Identify the best expression and key metrics, explain weaknesses, and suggest edits to guide the next generation.",
  "input_schema": {
    "H_struct": "<object>",
    "candidates": [
      {"expression": "<string>", "metrics": {"IC": "<float>", "IR": "<float>", "MDD": "<float>"}}
    ]
  },
  "output_schema": {
    "summary": {"best_expression": "<string>", "key_metrics": "<string>"},
    "feedback": ["<string>"],
    "suggested_edits": [{"action": "<tighten|relax|shift>", "detail": "<string>"}]
  }
}
\end{tcblisting}

\textbf{MarketLogicGeneratorAgent.}
\begin{tcblisting}{listing only, breakable, colback=white, colframe=black, boxrule=0.5pt, listing options={basicstyle=\ttfamily\footnotesize, breaklines=true}}
{
  "system": "Generate new market logic H = <C,B> based on historical logics and feedback. Ensure the logic reflects plausible market mechanisms and remains consistent with the structured C/B schema.",
  "instruction": "Use H_init_lib to seed the first-round logic; in later rounds use H_init_lib, H_current, H_hist, E_hist, and fb_hist to propose a new market logic grounded in existing market mechanisms and distinct from prior logics. Output a human-readable logic with explicit C and B semantics.",
  "input_schema": {
    "H_init_lib": ["<string>"],
    "H_current": "<string or null>",
    "H_hist": ["<string>"],
    "E_hist": ["<object>"],
    "fb_hist": ["<object>"],
    "round": "<int>"
  },
  "output_schema": {
    "logic_text": "<string>",
    "c_text": "<string>",
    "b_text": "<string>"
  }
}
\end{tcblisting}

\textbf{MarketLogicRefinementDirectionAgent.}
\begin{tcblisting}{listing only, breakable, colback=white, colframe=black, boxrule=0.5pt, listing options={basicstyle=\ttfamily\footnotesize, breaklines=true}}
{
  "system": "Analyze evidence and propose logic refinement directions. Aggregate factor performance across time, market conditions, and risk dimensions; identify logic components that are too broad, vague, or mismatched with market structure.",
  "instruction": "Analyze historical evidence and propose logic refinements. Summarize factor performance distribution across time, market conditions, and risk dimensions, and identify logic components that are too broad, vague, or mismatched with market structure.",
  "input_schema": {
    "H_current": "<string>",
    "H_hist": ["<string>"],
    "E_hist": ["<object>"],
    "fb_hist": ["<object>"]
  },
  "output_schema": {
    "refinement_actions": [
      {"action": "<tighten|relax|shift|reweight>", "target": "<C or B field>", "detail": "<string>"}
    ],
    "focus_variables": ["<string>"],
    "horizon_suggestion": "<string>",
    "rationale": "<string>"
  }
}
\end{tcblisting}

\end{document}